\documentstyle[aps,prl,epsfig,floats,preprint,tighten]{revtex}

\begin{document}

\draft

\date{September 2000}
\title {Single-mode delay time statistics for scattering by a chaotic cavity }
\author{K.~J.~H. van Bemmel, H. Schomerus, and C.~W.~J. Beenakker}
\address{Instituut-Lorentz, Universiteit Leiden, P.O. Box 9506, 2300 RA
Leiden, The Netherlands}

\widetext

\maketitle

\begin{abstract}
We investigate the low-frequency dynamics for transmission or 
reflection of a wave by a cavity with chaotic scattering. We compute the probability
distribution of the phase derivative $\phi '={\mathrm{d}}\phi /{\mathrm{d}}\omega$ of 
the scattered wave amplitude, known as the single-mode delay time. In the case of a
cavity connected to two single-mode waveguides we find a marked distinction between
detection in transmission and in reflection: The distribution $P(\phi ')$ vanishes for negative 
$\phi '$ in the first case but not in the second case.
\end{abstract}

\vspace{0.5cm}


\section{Introduction}

Microwave cavities have proven to be a good experimental testing ground for theories
of chaotic scattering\cite{boekStockmann}. Much work has been done on static scattering
properties, but recently dynamic features have been measured as well\cite{Persson}. A key dynamical
observable, introduced by Genack and coworkers\cite{Sebbah,Genack,Tiggelen}, is the
frequency derivative $\phi '={\mathrm{d}}\phi /{\mathrm{d}} \omega$ of the phase of 
the wave amplitude measured in a single speckle of the transmitted or reflected wave.
Because one speckle corresponds to one element of the scattering matrix, and because
$\phi '$ has the dimension of time, this quantity is called the single-channel or
single-mode delay time. It is a linear superposition of the Wigner-Smith delay times
introduced in nuclear physics\cite{Wigner,Smith}. 

The probability distribution of the
Wigner-Smith delay times for scattering by a chaotic cavity is
known\cite{properdelaytimes}. The purpose of this paper is to derive from that the
distribution $P(\phi ')$ of the single-mode delay time. The calculation follows closely
our previous calculation of $P(\phi ')$ for reflection from a disordered
waveguide in the localized regime\cite{ourpreprint}. The absence of localization in a chaotic cavity is a 
significant simplification. For a small number of modes $N$ connecting the cavity to the
outside we can calculate $P(\phi ')$ exactly, while for $N \gg 1 $ we can use 
perturbation theory in $1/N$. The large-$N$ distribution has the same form as that 
following from diffusion theory in a disordered waveguide\cite{Genack,Tiggelen}, 
but for small $N$ the distribution is qualitatively different. In particular, there
is a marked distinction between the distribution in transmission and in reflection.

\section{Formulation of the problem}

\begin{figure}[!tb]
\begin{center}
\includegraphics[angle=0, width=8cm]{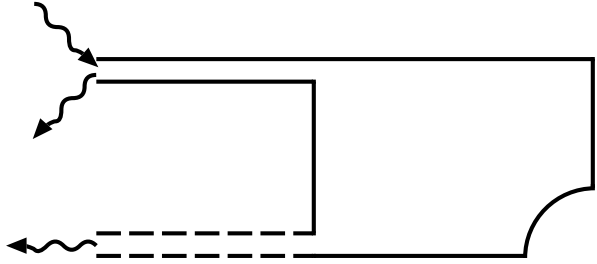}
\end{center}
\caption[x]{Sketch of a chaotic cavity coupled to $N$ propagating modes via one or
more waveguides. The shape of the cavity is the quartered Sinai billiard used in
recent microwave experiments\cite{Persson}.} \label {figSinai}
\end{figure}

The geometry studied is shown schematically in Figure \ref{figSinai}. It consists of an $N$-mode waveguide
connected at one end to a chaotic cavity. Reflections at the connection between
waveguide and cavity are neglected (ideal impedance matching). The $N$ modes may be
divided among different waveguides, for example, $N=2$ could refer to two single-mode
waveguides. The cavity may contain a ferri-magnetic element as in 
Refs.~\cite{So,Stoffregen}, in which
case time-reversal symmetry is broken. The symmetry index $\beta=1$ $(2)$ indicates the presence 
(absence) of time-reversal symmetry. We assume a single polarization for simplicity, as in the 
microwave experiments in a planar cavity\cite{Persson}.

The dynamical observable is the correlator $\rho $ of an element of the scattering 
matrix $S(\omega)$ at two nearby frequencies,

\begin{equation} \label {eq:corr}
\rho=S_{nm}(\omega +\case{1}{2} \delta \omega)S^*_{nm}(\omega -\case{1}{2} \delta \omega).
\end{equation}
The indices $n$ and $m$ indicate the detected outgoing mode and the incident mode, respectively. 
The single-mode delay time $\phi '$ is defined by \cite{Sebbah,Genack,Tiggelen}

\begin{equation} \label {eq:defdelaytime}
\phi ' = \lim_{\delta \omega \to 0} \frac{\mathrm{Im} \rho}{\delta \omega I},
\end{equation}
with $I=|S_{nm}(\omega)|^2$ the intensity of the scattered wave in 
mode $n$ for unit incident intensity in mode $m$. If we write the scattering 
amplitude $S_{nm}=\sqrt{I}{\mathrm{e}}^{i\phi}$ in terms of amplitude and phase,
 then $\phi '={\mathrm{d}}\phi /{\mathrm{d}}\omega$. We will investigate the  
distribution of $\phi '$ in an ensemble of chaotic cavities having
slightly different shape, at a given 
mean frequency interval $\Delta$ between the cavity modes. For notational
convenience, we choose units of time such that $2\pi/\Delta\equiv 1$.

The single-mode delay times are linearly related to the Wigner-Smith \cite{Wigner,Smith} delay
times $\tau_1,\tau_2,...,\tau_N$, which are the eigenvalues of the matrix
\begin{equation} \label {eq:Qmat}
Q=-iS^{\dagger}\frac{{\mathrm{d}}S}{{\mathrm{d}}\omega}=U^{\dagger}{\mathrm{diag}}(\tau_1,...,\tau_N)U.
\end{equation}
To see this, we first expand the scattering matrix linearly in $\delta \omega$,
\begin{equation} \label {eq:rexpansion}
S(\omega \pm \case{1}{2}\delta \omega)=V^{{\mathrm{T}}}U\pm \case{1}{2}i\delta \omega V^{{\mathrm{T}}}{\mathrm{diag}}(\tau_1,...,\tau_N)U.
\end{equation}
Since $S$ is symmetric for $\beta=1$, one then has $V=U$. For $\beta=2$, $V$ and $U$ are statistically
independent. Combination of Eqs.~($\ref{eq:corr}$), ($\ref{eq:defdelaytime}$), and ($\ref{eq:rexpansion}$) leads to \cite{ourpreprint}

\begin{eqnarray} \label {eq:expresphi}
&&I=|\sum_i^{} u_iv_i|^2, \hspace{1ex} \phi ' ={\mathrm{Re}} \hspace{1ex} \frac{\sum_{i}^{}\tau_i u_iv_i}{\sum_{j}^{} u_jv_j},
\\
&&u_i=U_{im}, \hspace{1ex} v_i=V_{in}.
\end{eqnarray} 

The distribution of the Wigner-Smith delay times for a chaotic cavity was calculated in Ref.~\cite{properdelaytimes}. It is a Laguerre ensemble in the rates $\mu_i=1/\tau_i$,
\begin{eqnarray} \label {eq:Lag}
P(\mu_1,...,\mu_N)\propto \prod_{i<j}^{}|\mu_i-\mu_j|^{\beta}\prod_k^{}\mu_k^{\beta N/2}\exp(-\case{1}{2}\beta \mu_k)\theta (\mu_k).
\end{eqnarray}
The step function $\theta (x)=1$ for $x>0$ and $\theta(x)=0$ for $x<0$. 
It follows from
Eq.~($\ref{eq:Lag}$) that $\langle \sum_i^{} \tau_i \rangle =1$, a result that was known
previously\cite{trace}.

To calculate the joint distribution $P(I,\phi ')$ from 
Eq.~($\ref{eq:expresphi}$), we also need the distribution of the coefficients $u_i$ 
and $v_i$. This follows from the Wigner conjecture\cite{conjecture}, proven in Ref.~\cite{properdelaytimes}, according to which the matrices $U$ and $V$ are uniformly distributed in the 
unitary group. The calculation for small $N$ is now a straightforward integration, see
Sec.~\ref{secsmallN}.  For large $N$ we can use perturbation theory, see Sec.~\ref{seclargeN}.

Because of the uniform distribution of $U$ and $V$, independent of the 
$\tau_i$'s, we can evaluate the average of $\phi '$ directly for any $N$,

\begin{equation}
\langle \phi ' \rangle ={\mathrm{Re}}  \left \langle \sum_i^{} \tau _i \left \langle \frac{u_i v_i}{\sum_j^{} u_j v_j} \right \rangle \right \rangle =\left \langle \sum_i^{} \tau_i \frac{1}{N} \right \rangle =\frac{1}{N}.
\end{equation}
We define the rescaled variable $\hat{\phi '}=\phi '/\langle \phi ' \rangle =N\phi '$,
that we will use in the next sections.

\section{Small number of modes} \label {secsmallN}
 
For $N=1$ there is no difference between 
the Wigner-Smith delay time and the single-mode delay time. In that case
$I=1$ and $\hat{\phi '}=\phi '$ is distributed according to \cite{Fyod-Som-Savin,Gopar}

\begin{equation} \label {eq:n1}
P(\hat{\phi '})=c_{\beta}\hat{\phi '}^{-2-\beta/2}\exp(-\case{1}{2}\beta/\hat{\phi '}) \theta (\hat{\phi '}).
\end{equation}
The normalization coefficient $c_{\beta}$ equals $(2\pi)^{-1/2}$ for $\beta =1$ and
$1$ for $\beta =2$.

Now we turn to the case $N=2$.
By writing out the summation in Eq.~($\ref{eq:expresphi}$) for $I$ and $\phi '$, one
obtains $\phi '=\tau_{+}+\alpha \tau_{-}$ with $\tau_{\pm}=\case{1}{2}(\tau_1 \pm \tau_2)$ and 

\begin{eqnarray} \label{eq:explIpsi}
&&I=|u_1|^2|v_1|^2+|u_2|^2|v_2|^2+u_1u_2^{*}v_1v_2^{*}+u_1^{*}u_2v_1^{*}v_2,
\\
\label{eq:explpsi}
&&\alpha =(|u_1|^2|v_1|^2-|u_2|^2|v_2|^2)/I.
\end{eqnarray}
To find the joint distribution $P(I,\alpha)$ we parametrize $U$ in terms
of $4$ independent angles,
 
\begin{equation} \label {eq:para}
U=
\left( \begin{array} {cc}
\cos \gamma \exp (-i\alpha _1) & \sin \gamma \exp (-i \alpha _1 -i \alpha _2) \\
-\sin \gamma \exp (-i \alpha _3 +i \alpha _2) & \cos \gamma \exp (-i\alpha _3) \\
\end{array} \right),
\end{equation}
with $\alpha _i\in(0,2\pi)$ and $\gamma\in(0,\pi/2)$.
The invariant measure ${\mathrm{d}}\mu \propto |{\mathrm{Det}} \hspace{1ex} g|{\mathrm{d}}\gamma
\prod _{i}^{} {\mathrm{d}} \alpha_i$ in the unitary group follows from
the metric tensor $g$, defined by
 
\begin{equation}
{\mathrm{Tr}} \hspace{1ex} {\mathrm{d}}U {\mathrm{d}} U^{\dagger}=\sum_{i,j}^{} g_{ij} {\mathrm{d}} x_i {\mathrm{d}} x_j, \hspace{1ex} \{x_i\}=\{\gamma,\alpha_1,\alpha_2,\alpha_3\}.
\end{equation}
The result is

\begin{equation} \label{eq:measure}
{\mathrm{d}}\mu \propto \sin 2\gamma \hspace{1ex} {\mathrm{d}}\gamma \prod _{i}^{} {\mathrm{d}} \alpha_i.
\end{equation}
The joint distribution function $P(\tau_+,\tau_-)$ follows from Eq.~($\ref{eq:Lag}$). For
$\beta =1$ one has

\begin{equation} \label{eq:tausbeta1}
P(\tau_{+},\tau_{-})=\case{1}{12}|\tau_{-}|(\tau_{+}^2-\tau_{-}^2)^{-4}\exp \left (-\tau_{+}(\tau_{+}^2-\tau_{-}^2)^{-1}\right)\theta (\tau_{+}-|\tau_{-}|),
\end{equation}
while for $\beta =2$

\begin{equation} \label {eq:tausbeta2}
P(\tau_{+},\tau_{-})=\case{1}{3}\tau_{-}^2(\tau_{+}^2-\tau_{-}^2)^{-6}\exp \left(-2\tau_{+}(\tau_{+}^2-\tau_{-}^2)^{-1}\right)\theta (\tau_{+}-|\tau_{-}|).
\end{equation}

First we consider the case $\beta=1$, $n\neq m$. Because of the unitarity of $U$, one has
$|v_1|^2=|u_2|^2$ and $|v_2|^2=|u_1|^2$. Therefore $\alpha =0$ and $\phi ' =\tau_{+}$, so
$\phi '$ is independent of $I$. Integration of Eq.~($\ref{eq:tausbeta1}$) over $\tau_{-}$ 
results in

\begin{equation} \label {eq:n2nonm}
P(\hat{\phi '})=\case{2}{3}\hat{\phi '}^{-5}(\hat{\phi '}^2+2\hat{\phi '}+2) \exp (-2/\hat{\phi '}) \theta (\hat{\phi '}).
\end{equation}
In this case (as well as in the case $N=1$), $\hat{\phi '}$ can take on only positive values, but this is atypical,
as we will see shortly. 
From Eqs.~($\ref{eq:explIpsi}$) and ($\ref{eq:para}$) we find the relation between $I$ and 
the parametrization of $U$,

\begin{equation}
I=\sin ^{2} 2\gamma \sin ^{2} (\alpha_3-\alpha_1-\alpha_2).
\end{equation}
The distribution of $I$ resulting from the measure ($\ref{eq:measure}$) is

\begin{equation} \label{eq:IN2trans}
P(I)=\case{1}{2} I^{-1/2} \theta (I) \theta (1-I),
\end{equation}
in agreement with Refs.~\cite{Baranger,Jalabert}.

\begin{figure}[!tb]
\begin{center}
\includegraphics[angle=0, width=8cm]{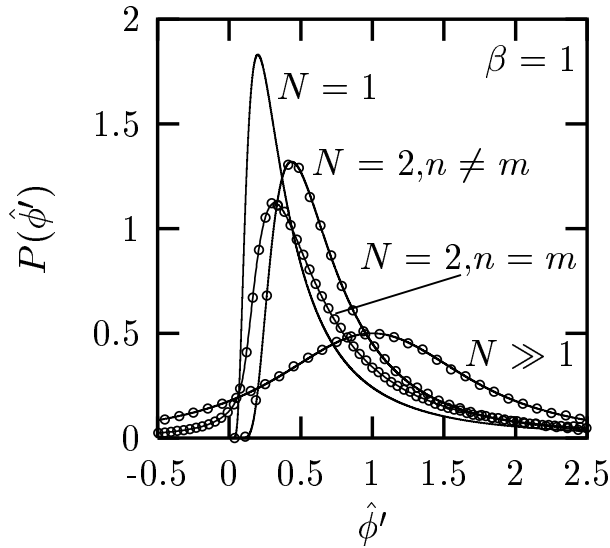}
\end{center}
\caption[x]{Distribution of the single-mode delay time in the case
of preserved time-reversal symmetry ($\beta =1$).
The curves for $N=1,2$ follow from Eqs.~($\ref{eq:n1}$), ($\ref{eq:n2nonm}$), and ($\ref{eq:n2rest}$).
The curve for $N\gg 1$ follows from Eq.~($\ref{eq:largeNphi}$), and is the same for $n=m$ and $n\neq m$. The delay time $\hat {\phi ' }=\phi '/\langle \phi ' \rangle$ is rescaled such that the mean is $1$ for all curves. Data
points are the result of a Monte Carlo calculation in the Laguerre ensemble
(with $N=400$, $n \neq m$ representing the large-$N$ limit).} \label {figcrossoverbeta1}
\end{figure}

For the case $N=2$, $\beta =1$, $n=m$ we use that $u_1=v_1$, $u_2=v_2$ and obtain 
the parametrization

\begin{eqnarray}
&&I=1-\sin ^2 2\gamma \sin ^2(\alpha_3-\alpha_1-\alpha_2),
\\
&&\alpha=(\cos 2\gamma)/I .
\end{eqnarray}
The distribution $P(I,\alpha)$ resulting from the measure ($\ref{eq:measure}$)
is
\begin{equation} \label {eq:fbeta1}
P(I,\alpha)=\frac{1}{2\pi} I^{1/2}(1-I)^{-1/2}(1-I\alpha^2)^{-1/2} \theta(I) \theta (1-I) \theta (1-I\alpha^2).
\end{equation}
The joint distribution of $I$ and $\hat {\phi '}=2\phi '$ takes the form 
\begin{equation} \label {eq:n2rest}
P(I,\hat {\phi '})=\int _{0}^{\infty} {\mathrm{d}}\tau_{-}\int_{\tau_{-}}^{\infty}
{\mathrm{d}}\tau_{+} \hspace{1ex} P(\tau_{+},\tau_{-})P\Big(I,\alpha=\frac{\hat {\phi '}/2-\tau_{+}}{\tau_{-}}\Big)\frac{1}{\tau_{-}}.
\end{equation}
The distribution of $I$ following from integration of $P(I,\alpha)$ over $\alpha$ is given
by Eq.~($\ref{eq:IN2trans}$) with $I \to 1-I$, as it should.
The integrations over $\tau_{+}, \tau_{-}$, and $I$, needed to obtain $P(\hat {\phi '})$ can be evaluated numerically, see Fig.~\ref{figcrossoverbeta1}.
Notice that $P(\hat {\phi '})$ has a tail towards negative values of $\hat{\phi '}$.

\begin{figure}[!tb]
\begin{center}
\includegraphics[angle=0, width=8cm]{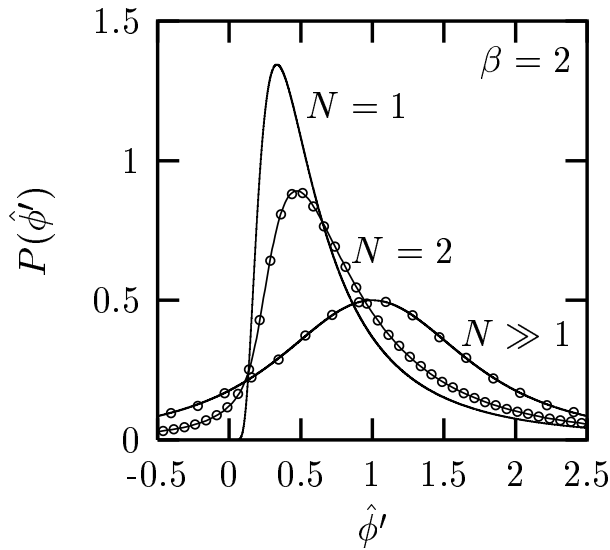}
\end{center}
\caption[x]{Same as in Fig.~\ref{figcrossoverbeta1}, for broken time-reversal
symmetry ($\beta=2$). The curves for $N=1,2$ and for $N\gg 1$ follow from
Eqs.~($\ref{eq:n1}$), ($\ref{eq:n2rest}$), and ($\ref{eq:largeNphi}$). There
is
no difference between $n=m$ and $n\neq m$ for any $N$. The large $N$-result
for $\beta=2$ is the same as for $\beta=1$.} \label {figcrossoverbeta2}
\end{figure}

For $N=2, \beta=2$ it doesn't matter whether $n$ and $m$ are equal or not.
Parametrization of both $U$ and $V$ leads to 

\begin{eqnarray}
&&I=(1-x_1)(1-x_2)+x_1x_2+2 \sqrt{(1-x_1)(1-x_2)x_1x_2} \cos \eta,
\\
&&\alpha=(1-x_1-x_2)/I,
\end{eqnarray}
with a measure ${\mathrm{d}}\mu \propto {\mathrm{d}}x_1 {\mathrm{d}} x_2 {\mathrm{d}} \eta$ and $x_1,x_2 \in(0,1)$, $\eta \in(0,\pi)$. The joint distribution $P(I,\alpha)$ is now given by 

\begin{equation} \label {eq:fbeta2}
P(I,\alpha)=\case{1}{2}I^{1/2} \theta (I) \theta (1-I) \theta (1-I\alpha^2).
\end{equation}
Integration over $\alpha$ leads to\cite{Baranger,Jalabert}

\begin{equation}
P(I)=\theta (I) \theta (1-I).
\end{equation}
The distribution $P(I,\hat {\phi '})$ follows upon insertion of Eqs.~($\ref{eq:tausbeta2}$) and ($\ref{eq:fbeta2}$) into Eq.~($\ref{eq:n2rest}$).
Numerical integration yields the distribution $P(\hat {\phi '})$ plotted in 
Fig.~\ref{figcrossoverbeta2}. As in the previous case, there is a tail
towards negative $\hat{\phi '}$.

\section{Large number of modes} \label {seclargeN}

We now calculate the joint distribution $P(I,\phi ')$ for $N\gg 1$.
First the case $n\neq m$ will be considered, when
there is no distinction between $\beta=1$ and $\beta=2$.
In the large $N$-limit
the vectors $\bf{u}$ and $\bf{v}$ become uncorrelated and their elements 
become independent Gaussian numbers with zero mean and variance $1/N$. 
We first average over $\bf{v}$, following Ref.~\cite{ourpreprint}.
We introduce the weighted delay 
time $W=I\phi '$. The Fourier transform of $P(I,W)$ is given by 
$\chi (p,q)=\langle \exp(ipI+iqW)\rangle $.
The average over $\bf{v}$ is a Gaussian integration, that gives 

\begin{eqnarray} \label{eq:det}
&&\chi (p,q)=\langle {\mathrm{det}}(1-iH/N)^{-1}\rangle ,
\\
&&H=p\textbf{u}^*\textbf{u}^{\mathrm{T}}+\case{1}{2}q(\bar{\textbf{u}}^*\textbf{u}^{\mathrm{T}}+\textbf{u}^*\bar{\textbf{u}}^{\mathrm{T}}),
\end{eqnarray}
where $\bar{u}_i=u_i\tau_i$. The matrix $H$ has only two nonzero eigenvalues,

\begin{eqnarray}
&&\lambda_{\pm}=\case{1}{2}\left(qB_1+p\pm\sqrt{2pqB_1+q^2B_2+p^2}\right),
\\
&&B_k=\sum_{i}^{}|u_i|^2\tau_i^k.
\end{eqnarray}
Performing the inverse Fourier transforms and returning to the variables $\phi '$ 
and $I$ leads to

\begin{equation} \label {eq:Iphi}
P(I,\phi ')=(N^3I/\pi)^{1/2}\exp(-NI)\left\langle (B_2-B_1^2)^{-1/2}\exp\left(-NI\frac{(\phi ' -B_1)^2}{B_2-B_1^2}\right)\right\rangle \theta (I).
\end{equation}
The averages over $u_i$ and $\tau_i$ still have to be performed.

Up to now the derivation is the same as for the disordered waveguide in the 
localized regime\cite{ourpreprint}, the only difference being the different distribution of
the Wigner-Smith delay times $\tau_i$. The absence of localization in a chaotic 
cavity greatly simplifies the subsequent calculation in our present case.
While in the localized waveguide anomalously large $\tau_i$'s lead to large 
fluctuations in $B_1$ and $B_2$, in the chaotic cavity the term $\mu _k ^{\beta N/2}$ in
Eq.~($\ref{eq:Lag}$) suppresses large delay times. Fluctuations in $B_k$ are smaller
than the mean by a factor $1/\sqrt{N}$. For $N \gg 1$ we may therefore replace $B_k$ in Eq.~($\ref{eq:Iphi}$) by $\langle B_k \rangle$.

To calculate the average of $B_1$ and $B_2$ we need the density $\rho (\tau)=\langle
 \sum _i ^{} \delta (\tau -\tau_i)\rangle $ of the delay times. It is given by
\cite{properdelaytimes}

\begin{equation} \label {eq:rhotau}
\rho (\tau)=\frac{N}{2\pi\tau ^2}\sqrt {(\tau_{+}-\tau)(\tau-\tau_{-})}, \hspace{1ex} \tau_{\pm}=\frac{3\pm \sqrt{8}}{N},
\end{equation}
for $\tau$ inside the interval $(\tau_{-},\tau_{+})$. The density is zero outside this interval.
The resulting averages are $\langle B_1 \rangle =N^{-1}$ and $\langle B_2\rangle =2N^{-2}$,
which leads to

\begin{equation} \label {eq:nmjoint}
P(I,\hat{\phi '})= (N^3I/\pi)^{1/2}\exp\left(-NI \left[1+(\hat{\phi '}-1)^2\right]\right) \theta (I).
\end{equation}
(Recall that $\hat {\phi '}=\phi '/\langle \phi ' \rangle =N\phi '$.)
Integration over $\hat{\phi '}$ or $I$ gives

\begin{eqnarray} 
P(I)&=&N\exp(-NI) \theta (I),
\\
\label{eq:largeNphi}
P(\hat{\phi '})&=&\case{1}{2}\left[1+(\hat{\phi '}-1)^2\right]^{-3/2}.
\end{eqnarray}
This distribution of $I$ and $\hat {\phi '}$ has the same form as that of a disordered waveguide in the 
diffusive regime \cite{Genack,Tiggelen}.

We next turn to the case $n=m$ and $\beta=1$. (For $\beta=2$ there is no
difference between $n=m$ and $n\neq m$.)
Since $u_i=v_i$ in Eq.~($\ref{eq:expresphi}$) we have

\begin{equation} \label {eq:phinn}
I=|C_0|^2, \hspace{1ex} \phi '={\mathrm{Re}} \hspace{1ex} \frac{C_1}{C_0}, \hspace{1ex} C_k=\sum_i^{} \tau_i^k u_i^2.
\end{equation}
The joint distribution $P(C_0,C_1)$ has the Fourier transform

\begin{equation}
\chi (p_0,p_1,q_0,q_1)=\langle\exp(ip_0{\mathrm{Re}}C_0+iq_0{\mathrm{Im}}C_0+ip_1{\mathrm{Re}}C_1+iq_1{\mathrm{Im}}C_1)\rangle.
\end{equation}
Averaging over $\bf{u}$ we find  

\begin{eqnarray}
&&\chi (p_0,p_1,q_0,q_1)=\langle \exp (-L) \rangle,
\\
\label{eq:LbigNnn}
&&L=\case{1}{2} \sum _i^{}  \ln \left[1+N^{-2} (p_0+p_1\tau_i)^2+N^{-2}(q_0+q_1\tau_i)^2\right].
\end{eqnarray}
Fluctuations in $L$ are smaller than the average by a factor $1/N$. We may therefore
approximate $\langle \exp (-L) \rangle \approx \exp \langle -L \rangle$. Because $N^{-2} (p_0+p_1\tau_i)^2+N^{-2}(q_0+q_1\tau_i)^2$ is of order $1/N$, we may expand the 
logarithm in Eq.~($\ref{eq:LbigNnn}$). The average follows upon integration with the density ($\ref{eq:rhotau}$),

\begin{equation}
\langle L \rangle =\frac{p_0^2+q_0^2}{2N}+\frac{p_1^2+q_1^2}{N^3}+\frac{p_0p_1+q_0q_1}{N^2}.
\end{equation}
Inverse Fourier transformation gives

\begin{equation}
P(C_0,C_1)=\frac{N^4}{(2\pi)^2}\exp (-N|C_0|^2-\case{1}{2}N^3|C_1|^2+N^2{\mathrm{Re}} \hspace{1ex} C_0C_1^*).
\end{equation}
The resulting distribution of $\hat {\phi '}$ and $I$ is

\begin{equation} \label {eq:largeNnm}
P(I,\hat{\phi '})=(N^3I/2\pi)^{1/2} \exp \left(-\case{1}{2}NI\left[1+(\hat{\phi '}-1)^2\right]\right)\theta (I).
\end{equation}
It is the same as the distribution ($\ref{eq:nmjoint}$) for $n\neq m$, apart from the rescaling of
$I$ by a factor of $2$ as a result of coherent backscattering.

The distribution ($\ref{eq:largeNphi}$) of $\hat {\phi '}$ for $N \gg 1 $ is included in Figs.~\ref{figcrossoverbeta1} and \ref{figcrossoverbeta2} for comparison with the small $N$-results.

\section{Numerical check}

We can check our analytical calculations by performing a Monte Carlo average over
the Laguerre ensemble for the $\tau_i$'s and the unitary group for the $u_i$'s
and $v_i$'s. For the average over the unitary group we generate a large number of
complex Hermitian $N\times N$ matrices $H$. The real and imaginary 
part of the off-diagonal elements are independently Gaussian distributed with 
zero mean and unit variance. The real diagonal elements are 
independently Gaussian distributed with
zero mean and variance 2. We diagonalize $H$, order the eigenvalues from large to
small, and multiply the $n$-th normalized eigenvector 
by a random phase factor ${\mathrm{e}}^{i\alpha _n}$, with $\alpha _n$ 
chosen uniformly from $(0,2\pi)$. The resulting matrix of eigenvectors is 
uniformly distributed in the unitary group.

The Laguerre ensemble ($\ref{eq:Lag}$) for the rates $\mu_i=1/\tau_i$ can be 
generated by a random matrix of the Wishart type \cite{Edelman,Baker}. Consider
a $N\times (2N-1+2/\beta)$ matrix $X$, where $X$ is real for $\beta =1$ and complex for 
$\beta =2$. (The matrix $X$ is neither symmetric nor Hermitian.) The matrix elements are 
Gaussian distributed with zero mean and variance $\langle |x_{nm}|^2\rangle =1$. 
The joint probability distribution of the eigenvalues of the 
matrix $XX^{\dagger}$ is then given by Eq.~($\ref{eq:Lag}$). The results of our 
numerical check are included in Figs.~\ref{figcrossoverbeta1} and \ref{figcrossoverbeta2}. 
The large-$N$ limit is represented by $N=400$, $n \neq m$.
The analytical curves agree well with the numerical data.

\section{Conclusion}

We have investigated the statistics of the single-mode delay time $\phi '$ for chaotic
scattering. For a large number $N$ of scattering channels the distribution
has the same form as for diffusive scattering\cite{Genack,Tiggelen}, but for 
small $N$ the distribution is different. The case $N=2$ is of particular interest, 
representing a cavity connected to two single-mode waveguides. For preserved time-reversal
symmetry and detection in transmission ($\beta =1, n\neq m$), we find that $\phi '$ 
can take on only positive values, similarly to the 
Wigner-Smith delay times. In contrast, for detection in reflection (or for broken
time-reversal symmetry) the distribution acquires a tail towards negative $\phi '$.
These theoretical predictions are amenable to experimental test in the microwave 
cavities of current interest\cite{Persson}.

\vspace{2ex}

\noindent{\bf{Acknowledgements}}

\vspace{1ex}

We thank P.W.~Brouwer and M.~Patra for valuable advice. This research was supported by the 
``Nederlandse organisatie voor Wetenschappelijk Onderzoek'' (NWO) and by the ``Stichting
voor Fundamenteel Onderzoek der Materie'' (FOM).

\end{document}